\documentclass[twocolumn,prl]{revtex4}

\usepackage{graphicx}      
\usepackage{amsmath}       
\usepackage{amssymb}       
\usepackage{bm}            
\usepackage{hyperref}      
\hypersetup{colorlinks,citecolor=black,filecolor=black,linkcolor=black,urlcolor=black}

\begin{document}

\title{The Energy Landscape of Social Balance}

\author{Seth A. Marvel\footnotemark[2]}

\author{Steven H. Strogatz\footnotemark[2]}

\author{Jon M. Kleinberg\footnotemark[3]}
\email{kleinber@cs.cornell.edu}

\affiliation{\vspace{6 pt} \footnotemark[2]Center for Applied Mathematics, Cornell University, Ithaca, New York 14853 \vspace{1 pt} \\
\footnotemark[3]Department of Computer Science, Cornell University, Ithaca, New York 14853}

\begin{abstract}

We model a close-knit community of friends and enemies as a fully connected network with positive and negative signs on its edges.  Theories from social psychology suggest that certain sign patterns are more stable than others.  This notion of social ``balance'' allows us to define an energy landscape for such networks.  Its structure is complex: numerical experiments reveal a landscape dimpled with local minima of widely varying energy levels.  We derive rigorous bounds on the energies of these local minima and prove that they have a modular structure that can be used to classify them.

\end{abstract}

\maketitle

The shifting of alliances and rivalries in a social group 
can be viewed as arising from an energy minimization process.
For example, suppose you have two friends who happen to detest each other.
The resulting awkwardness often resolves
itself in one of two ways: either you drop one of your friends, or 
they find a way to reconcile.  In such
scenarios, the overall social stress corresponds to a kind of
energy that relaxes over time as relationships switch from hostility to friendship or vice versa.

This view, now known as balance theory, was first articulated by 
Heider~\cite{heid46,heid58} and has since been applied in fields ranging from anthropology to political science~\cite{taylor70, moor79}.  
In the 1950s, Cartwright and Harary
converted Heider's conceptual framework to a graph-theoretic model and
characterized the global minima of the social energy
landscape~\cite{cart56}.  Their tidy analysis gave no hint that the energy landscape 
was anything more complicated than a series of equally deep wells, each 
achieving the minimum possible energy.  Recently, however, Antal,
Krapivsky and Redner~\cite{anta05} observed that the energy landscape also contains local minima, which they called \textit{jammed states}.

Jammed states are important to understand because they can trap a system as it moves down the energy landscape.  Yet little is known about their allowed energies, their structure, or how they depend on the size of the network.  
Even the maximum possible
energy of a jammed state is not obvious:
a simple argument (see below) shows that jammed states cannot be 
located more than halfway up the energy spectrum, but it is hard to see whether this upper bound can be achieved.

In this Letter we prove that for arbitrarily large networks, there do indeed exist jammed states all the way up to the midpoint energy, using a construction based on highly symmetric structures first discovered by Paley in
his work on orthogonal matrices~\cite{pale33}.  We also show
 that jammed states have a natural
modular structure.  This allows us to organize the jammed states encountered by simulation and to explain why
high-energy jammed states must be structurally more complex
than low-energy ones.

More broadly, our work here is part of a growing line of research
that employs tools from physics to analyze models of complex
social systems 
\cite{albert-revmodphys, dorogovtsev-net-book, newman-sirev}.
Theories of signed social
networks form an appealing domain for such techniques, as they are naturally cast
in the framework of energy minimization.

We begin by modeling a fully connected social network as a signed
complete graph on $n$ nodes. Each edge $\{i, j\}$ of the network
is labeled with either a plus or minus sign, denoted by $s_{ij}$,
corresponding to feelings of friendship or animosity
between the nodes $i$ and $j$.

	\begin{figure}[b]
	\includegraphics[scale = 1]{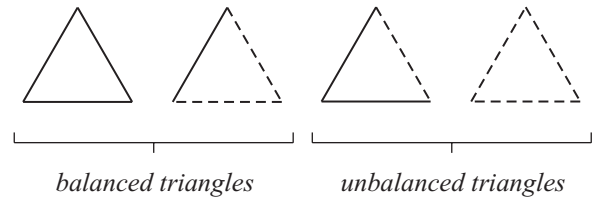}
	\caption{Socially balanced and unbalanced configurations of a triangle.  Solid edges represent friendly $(+)$ relationships, and dashed edges hostile $(-)$ relationships.  
  \label{fig1}}
	\end{figure}

Up to node permutation, there are four possible signings of a
triangle (Fig.~\ref{fig1}).  We view the two triangles
with an odd number of plus edges as balanced configurations,
since both satisfy the adages that ``the enemy of my enemy is my
friend,'' ``the friend of my
enemy is my enemy,'' and so on.
Since the two triangles with an even number of plus edges
break with this logic of friendship, we consider them unbalanced.

	\begin{figure}[b]
	\includegraphics[scale = 1]{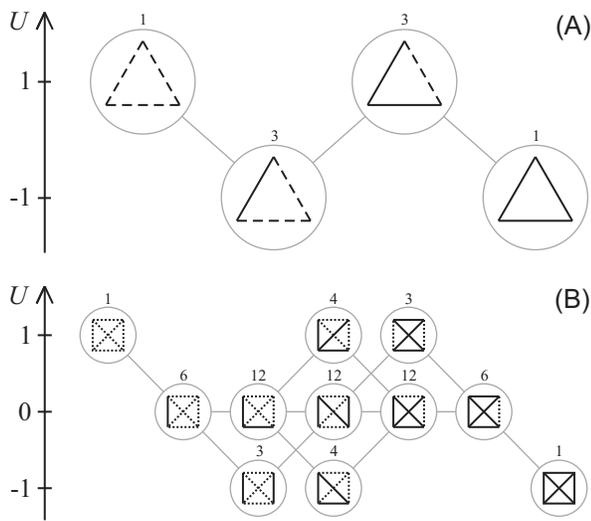}
	\caption{The energy landscapes of signed complete social networks on (A)~3 and (B)~4 nodes.  For simplicity, each set of sign configurations identical up to node permutation is represented by a single configuration; the number above each configuration indicates its multiplicity.  Lines between circles join networks differing by a single sign flip.  No jammed states occur for these small networks; they appear only when $n = 6$ or $n \ge 8$.  Strict jammed states occur when $n = 9$ and $n \ge 11$~\cite{anta05}.  \label{fig2}}
	\end{figure}

The product of the edge signs is positive for a balanced triangle and negative for an unbalanced triangle.  If we
sum the negative of these products and divide by the total number of
triangles, we obtain a quantity $U$ that represents the
elevation, or potential energy, of a social network above the
domain of all its possible sign configurations (Fig.~\ref{fig2}).
Explicitly,
 	\begin{equation} \label{1}
	U = -\frac{1}{\binom{n}{3}} \sum s_{ij} s_{jk} s_{ik}
	\end{equation}
where the sum is over all triangles $\{i, j, k\}$ of the network.  

The configuration in which all node pairs are friends has the lowest
possible energy $U = -1$.  Hence, no additional structure is necessary
to define the global minima; they are just the sign configurations for
which $U = -1$.  Cartwright and Harary~\cite{cart56} identified all such ground
states, finding that they consist of two warring factions: internally
friendly cliques with only antagonistic edges between them. (The all-friends
configuration represents the extreme in which one clique is the empty set.)

To define the concept of a local minimum, however, we need to specify 
what it means for two states to be adjacent.
The most natural choice is to define
two sign configurations to be adjacent if each can be reached from the
other by a single sign flip.  Then a jammed state, as defined by Antal
et al., is a sign configuration for which all adjacent sign
configurations have higher energy~\cite{anta05}.  Here,
however, we will slightly vary their terminology by calling this a
`strict jammed state,' reserving the term `jammed state' for the
weaker concept of a sign configuration with no adjacent sign
configurations of lower energy.

Our first result is that
jammed states cannot have energies above
zero.  To see this, note that every edge of a jammed state takes part in at least as many
balanced triangles as unbalanced triangles.
It is therefore found in $(n-2)/2$ unbalanced triangles if $n$ is
even and $(n-3)/2$ unbalanced triangles if $n$ is odd.
Thus, summing over all edges and dividing by 3 to avoid triple counting yields
 $U \leq -\frac{1}{3} \binom{n}{2} \bigl[\bigl(n-2 -
\frac{n-2}{2}\bigl) - \frac{n-2}{2}\bigl] / \binom{n}{3} = 0$ if $n$
is even and $U \leq -(n-2)^{-1}$ if $n$ is odd.

Are there jammed states that achieve this upper bound on $U$?
One possible way to address this question is through computational searches.
For example, suppose that from a random initial configuration,
we select and switch single signs uniformly at
random from the set of unbalanced edges (an edge is defined as
unbalanced if more than half the triangles that include it are
unbalanced).  We continue switching signs until the network reaches a
local minimum of $U$.  
Extensive searches of this form reveal only two small examples
of zero-energy jammed states:
a configuration on 6 nodes, consisting of a 5-cycle of positive edges
and all other edges negative, and 
a more complex configuration on 10 nodes.
Even on 10 nodes, only about 7 in $10^8$ searches end up at zero-energy
jammed states, and no such states were found on larger numbers of nodes.
The failure of this approach to produce even moderately-sized 
examples is consistent with findings of Antal et al. \cite{anta05},
who showed that such local search methods reach jammed states with a
probability that decreases to $0$ extremely rapidly as a function of
the network size $n$.

With only these data, the chances of finding a larger collection
of jammed states at $U = 0$ may seem slim.  
However, we now show how an infinite collection of zero-energy jammed states
can be identified through a direct combinatorial construction.
This construction is motivated by the two small examples 
found through computational searches:
when we re-examined the zero-energy jammed states on 6 and 10 nodes, 
we noticed that the positive edges formed so-called Paley graphs \cite{boll01}
on 5 and 9 nodes.  This beautiful connection turns out to be
general:  a family of arbitrarily large jammed states with $U = 0$ may be
derived from the undirected Paley graphs.

Briefly, an undirected Paley graph $P_q$ can be constructed on a set of $q$ nodes, where $q$ is a prime of the form $q = 4k + 1$ for some integer $k$.  To do so, we 
index the nodes with the integers $0, \dotsc, q - 1$ and then connect each
$v$ and $w$ in this node set with an edge if there is an $x$ in $\{0,
\dotsc, q-1\}$ such that $(v - w) \, \text{mod} \, q = x^2 \,
\text{mod} \, q$.  To construct the jammed state with $U = 0$ from $P_q$,
we give plus signs to the edges of $P_q$ and minus signs to the edges of its complement.  We then add a node $v_n$, where $n = q + 1$, and link it to all nodes of $P_q$ with negative edges.  (Paley graphs also exist if $q$ is a prime power, but then one needs to work over the finite field of order $q$.)

We now show that this new signed complete graph has zero energy.
Clearly, this is equivalent to the condition that each edge is in
exactly $\frac{n-2}{2}$ balanced triangles.  To check the latter
claim, we make use of two known properties of Paley graphs:  (i) $P_q$
is $2k$-regular, and (ii) for any two nodes $v$ and $w$ of $P_q$, there
are $k$ nodes adjacent to $v$ but not $w$, and $k$ nodes adjacent to
$w$ but not $v$~\cite{boll01}.

Now, if $\{v,w\}$ is a negative edge in $P_q$, then it forms balanced
triangles with all nodes $x$ in $P_q$ that are linked by a positive edge to
exactly one of $v$ or $w$.  By property (ii), there are $2k =
\frac{q-1}{2} = \frac{n-2}{2}$ such nodes, so $\{v,w\}$ is in exactly
$\frac{n-2}{2}$ balanced triangles.  Similarly, if $\{v,w\}$ is a positive 
edge in $P_q$, then it forms unbalanced triangles with all nodes $x$
of $P_q$ that are linked via a positive edge to exactly one of $v$ or $w$.
Again, these nodes account for $2k = \frac{n-2}{2}$ unbalanced
triangles, so $\{v,w\}$ is in exactly $\frac{n-2}{2}$ balanced
triangles.  Finally, since $P_q$ is $2k$-regular, there are exactly
$2k$ nodes in $P_q$ adjacent via positive edges to each node $w$ of $P_q$.
Hence, each negative edge $\{v_n,w\}$ is also in exactly $\frac{n-2}{2}$
balanced triangles.

The above construction is related to a result by Seidel regarding
two-graphs~\cite{seid76}.  
Using the theory of two-graphs, one can also construct 
infinite families of strict jammed states that approach $U = 0$ 
from below as $n$ grows large.  
Such constructions can be carried out using
bilinear forms modulo 2 \cite{seid76},
and projective planes in finite vector spaces
\cite{tayl71}.

Given the conceptual complexity of these constructions
of high-energy jammed states, and the computational
difficulty in finding such states via search,
it is natural to ask why it is harder to construct jammed
states closer to $U = 0$ than at lower energies.  
We now explain this by formulating a measure
of the complexity of different jammed states.
This will establish a precise sense in which higher-energy jammed states
are structurally more complex than lower-energy jammed states,
through a result showing that every signed complete graph has a natural
decomposition into internally balanced modules.

The statement of this \textit{edge balance decomposition} is
as follows.
Consider the subgraph $K$ consisting of all nodes in the network,
together with those edges that appear only in balanced triangles.
Then (i) $K$ is a union of disjoint cliques $\{C_a\}$ (possibly including
single-node cliques), and 
(ii) for every pair of cliques $C_a$ and $C_b$, every
edge between $C_a$ and $C_b$ is involved in the same number of
balanced triangles.  In the spirit of (i), we call each clique of the
partition a \textit{balanced clique}.

To prove part (i) of the decomposition, one can show that if some connected
component of $K$ is not a clique, then this component contains 
edges $\{i,j\}$ and $\{i,k\}$ sharing a node $i$ 
that are both found only in balanced triangles,
and such that $\{j,k\}$ is in at least one unbalanced triangle
(involving a fourth node $\ell$).
But then the set of four nodes $\{i, j, k, \ell\}$ would have three
of its four triangles balanced, which is not possible for any
sign pattern.

To prove part (ii) of the decomposition, one can show that if there were
cliques $C_a$ and $C_b$ such that two different edges between them
were involved in different numbers of unbalanced triangles,
then there would be two such edges $\{i,j\}$ and $\{i,k\}$ sharing
a node $i$ in $C_a$, such that for some other node $\ell$,
the triangle $\{i, j, \ell\}$ is balanced but the triangle $\{i, k, \ell\}$ is not.
But since $\{j,k\}$ is inside the clique $C_b$, all the triangles 
involving it are balanced, and so the four-node set 
$\{i, j, k, \ell\}$ would have three
of its four triangles balanced, which again is not possible for any
sign pattern.

We now return to the question that we posed above:  why is it harder
to construct jammed states near $U = 0$ than at substantially lower
energies?  We can close in on an elementary answer by computing an
upper bound on the allowed energy of a jammed state as a function of
the number of balanced cliques it contains.
We find that as the energy approaches $U = 0$ from below,
the number of cliques in the decomposition must grow unboundedly
in $n$, the number of nodes in the network.

First observe that for a fixed number of balanced cliques $m$, the
fewest number of edges are in balanced cliques---and hence the most
edges are available for inclusion in unbalanced triangles---when the
$n$ nodes of the network are equally distributed among the $m$
balanced cliques.  We can verify this using Lagrange
multipliers:  we seek to minimize $\sum_i \binom{c_i}{2}$ relative to
the constraints $\sum_i c_i = n$, $c_i > 0$, where $c_i$ is the number
of nodes in the $i$th balanced clique.  This implies $\frac{d}{dc_i}
\binom{c_i}{2} = \lambda$ for all $c_i$, where $\lambda$ is some
constant.  The derivative of the gamma function extension of
$\binom{c_i}{2}$ is monotone increasing on $c_i > 0$, so we invert it
to find all $c_i$ equal to the same function of $\lambda$.

Hence, no jammed state with $n$ nodes and $m$ balanced cliques can have greater energy than one in which the nodes are equidistributed among the balanced cliques and each edge spanning two balanced cliques participates in $\frac{n-2}{2}$ unbalanced triangles.  This implies an upper bound on $U$ of
 	\begin{equation} \label{2}
	U_n^{UB}(m) = -1 + 2 \frac{\frac{1}{3} \binom{m}{2} (\frac{n}{m})^2 \frac{n-2}{2}}{\binom{n}{3}} = -\frac{n-m}{m(n-1)}
	\end{equation}
For example, $\lim_{n \rightarrow \infty} U_n^{UB}(3) = -1/3$, whereas the corresponding tight upper bound (also verified by Lagrange multipliers) is $\lim_{n \rightarrow \infty} U = -\lim_{n \rightarrow \infty} [(\binom{n}{2} - (\frac{n}{3})^3) - (\frac{n}{3})^3] / \binom{n}{3} = -5/9$.

	\begin{figure}[t]
	\includegraphics[scale = 1]{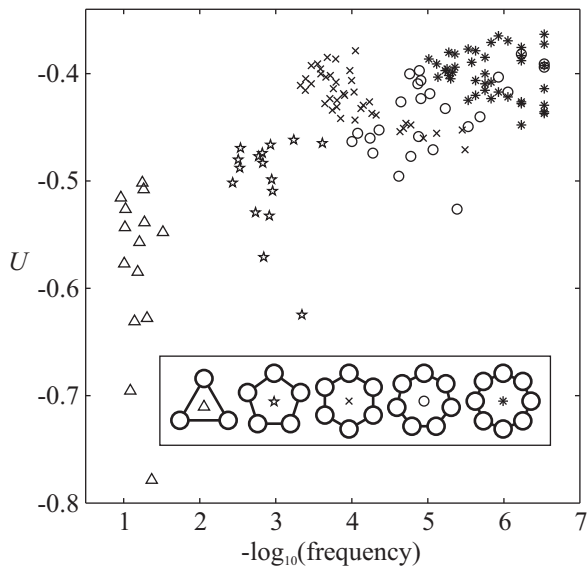}
\caption{Jammed states for networks with $n=26$ nodes, distinguished
according to their energy, frequency of occurrence, and clique
structure.  The different data symbols show the number of balanced
cliques in a given state (see inset for key).  We find that jammed
states with higher energies are not only rarer (as shown by
Antal et al. \cite{anta05})---they also have inherently greater
structural complexity, as measured by their number of balanced cliques.
To find these states, we evolved $10^8$ social networks to energy
minima via the Markov process described in the text, assuming that
each edge was initially unfriendly.  For simplicity, only
jammed states with eight or fewer balanced cliques are shown (these
comprised $> 99.99\%$ of all jammed states found).  Jammed states with
two and four balanced cliques are impossible.  Analogous distributions
for other $n$ and other initial sign patterns are similar, and increasing the number of
trial networks does not significantly change the distribution.  \label{fig4}}

	\end{figure}

We can see directly from (\ref{2}) that as we approach $U = 0$ from below, jammed states with $n$ nodes and $m$ or fewer balanced cliques no longer appear above $U_n^{UB}(m)$.  In other words, jammed states disappear as the energy is raised in order of least to greatest complexity.  Finally, at $U = 0$, the condition $U_n^{UB}(m) = 0$ implies that $m = n$, as we would expect since every edge must be in exactly $\frac{n-2}{2}$ balanced triangles.

In addition to illuminating a fundamental progression within the energy
spectrum of the jammed states, the edge balance decomposition also
provides a partition of the set of $2^{\binom{n}{2}}$ possible sign
configurations which proves useful for classifying jammed states.
Consistent with Antal et al.~\cite{anta05}, our numerical simulations of small
networks (generally $n < 2^{10}$) turned up an enormous number of
three-clique jammed states.  Less frequently, we encountered jammed
states with five, six and seven cliques, and rarely did we find jammed
states with more than seven cliques (Fig.~\ref{fig4}).  This numerical
evidence leads us to suspect that the most common jammed states found
in sign patterns arising from local search have only a few balanced
cliques and hence would be easily classified by the edge balance
decomposition.  (That said, it is possible to \textit{construct}
strict jammed states with $m$ balanced cliques for all odd $m$ in the
large-$n$ limit; whether such a
construction exists for even $m > 6$ remains open.)

In future work, it could be interesting to explore the model above
using tools from other parts of physics 
\cite{hertz, dedom80, wegner71, filk00, franz01}.
For example, the social balance model may be viewed as a generalized Ising model  \cite{wegner71} or $Z_2$ gauge theory \cite{filk00}  on the complete graph.  It is also similar to spin-glass models~\cite{hertz, dedom80}  where nodes in a network are likewise joined by edges of mixed signs, and $U$ measures the average frustration of the
system.  This line of work includes results on spin-glass systems with
three-way interactions \cite{franz01}, such as occur in Eq.~(\ref{1}).  One
potential obstacle to making this link is that in spin-glass models,
adjacency between configurations is defined  by changes in the signs
of nodes (due to spin flips) while edge signs remain fixed; whereas
here it is the signs of edges that vary as one moves across the
landscape.   This could possibly be addressed using transformations
that interchange the roles of nodes and edges; however, when the
complete graph is transformed in this way, the resulting network has a
complex structure that may hinder analysis.

Taken together, the results presented here yield a first look at the energy
spectrum of jammed states in completely connected social networks in
which opportunities for greater relational consistency and cooperation
are the driving forces for change.  
Since balanced states---the global energy minima---correspond 
to two antagonistic cliques, they often represent socially
undesirable outcomes such as intractable conflict.
Viewed in this light, the presence of jammed states at higher
energies suggests the possible beginnings of a companion theory of
reconciliation and flexibility in the setting of 
social balance: since these jammed states show less large-scale
antagonism, they may provide pathways to steer conflicts
into structures where reconciliation can more easily occur.

\vspace{4 pt}

\textbf{Acknowledgments:} 
Research supported in part by 
        the John D. and Catherine T.  MacArthur Foundation,
        a Google Research Grant,
        a Yahoo!~Research Alliance Grant,
        and NSF grants
        CCF-0325453, BCS-0537606, IIS-0705774, and CISE-0835706.

\end{document}